\documentstyle[prd,aps]{revtex}

\def\be{\begin{equation}}
\def\ee{\end{equation}}

\def\bea{\begin{eqnarray}}
\def\eea{\end{eqnarray}}
\def\bml{\begin{mathletters}}
\def\blea{\begin{mathletters}\begin{eqnarray}}
\def\elea{\end{eqnarray}\end{mathletters}}

\def\Tr{\mathop{\rm Tr}}

\def\ba{{\bf a}}
\def\bb{{\bf b}}
\def\bx{{\bf x}}

\def\xdot{{\dot\bx}}

\begin{document}
\draft
\wideabs{
\title{Dynamics of superconducting strings with chiral currents}

\author{J.\ J.\ Blanco-Pillado\footnote{Email address: {\tt
jose@cosmos.phy.tufts.edu}}, Ken D.\ Olum\footnote{Email address:
 {\tt kdo@alum.mit.edu}} and Alexander Vilenkin\footnote{Email address:{\tt 
vilenkin@cosmos.phy.tufts.edu}}}

\address{Institute of Cosmology, 
Department of Physics and Astronomy, 
Tufts University, 
Medford, Massachusetts 02155}

\date{April 2000; Revised December 2000}

\maketitle

\begin{abstract}%
We rederive, using an elementary formalism, the general solution to
the equations of motion for a superconducting string with a chiral
(null) neutral current, earlier obtained by Carter and Peter. We apply
this solution to show that the motion of such string loops is strictly
periodic and analyze cusp-like behavior and vorton solutions of
arbitrary shape. We argue that this solution can be used to
approximately describe the dynamics of superconducting strings with
small non-chiral currents. We use this description to estimate the
electromagnetic radiation power from such strings.
\end{abstract}

\pacs{98.80.Cq	
	11.27.+d 
}

}
\def\thefootnote{\fnsymbol{footnote}}
\footnotetext[1]{Email address: {\tt jose@cosmos.phy.tufts.edu}}
\footnotetext[2]{Email address: {\tt kdo@alum.mit.edu}}
\footnotetext[3]{Email address: {\tt vilenkin@cosmos.phy.tufts.edu}}
\def\thefootnote{\arabic{footnote}}
\narrowtext

\section{Introduction}
Cosmic strings are linear topological defects that may have been
created in the early Universe\cite{Kibble76}. They have been
extensively studied in connection with several problems in cosmology (for
reviews see\cite{Alexbook,Kibble95}).  Assuming that the radius of
curvature of a string is always much larger than the string core
thickness, the string dynamics can be described by the
Nambu-Goto action\cite{Nambu70,Goto71}. In this approximation the 
equations of motion are simple, and can be easily solved.

In 1985, Witten\cite{Witten85} showed that strings could behave as
superconducting wires in certain particle physics models. This new
internal degree of freedom opened up a variety of interesting
effects. In particular, superconducting strings may have stable
configurations, vortons\cite{Davis88-2} and
springs\cite{Haws88,Copeland88}, which could contribute to the dark
matter in the universe, or put constraints on the particle physics
models that give rise to those strings\cite{Carter:1999wy,Carter99-2}.
Superconducting cosmic strings have also been considered as sources
for structure formation\cite{Ostriker86}, gamma ray
bursts\cite{Babul87,Paczynski88,Berezinsky:2001cp}, and ultra-high energy
cosmic rays\cite{Berezinsky:2001cp,Hill87,Berezinsky98}.

The general problem of a superconducting string coupled to the
electromagnetic field cannot be solved analytically.  However, if the
charge carriers are not coupled to any long-range field (so-called
neutral superconducting strings)\cite{Shellard88} the situation is
significantly simpler.  Such strings are of interest in their own
right, and may also be considered as approximations to the full theory
including electromagnetic coupling.

If in addition, we consider the case that the charge and current are
equal in magnitude, then it was shown by Carter and Peter\cite{Carter99-1}
that the equations of motion can be solved
exactly.  In this
case, the charge-current 4-vector is lightlike, and the current is
said to be chiral or null.  It consists of charge carriers which are
all moving in the same direction.  Such currents arise automatically
in certain supersymmetric theories in which a zero mode can travel
only in one direction along the string \cite{Davis97,Carter99-2}.
They could also result from evolution of a loop with an arbitrary
distribution of charge and current\cite{Spergel87,Davis88-2,Martins98}.
Left- and right-moving charge carriers can scatter off the 
string, and if the numbers of left- and right-movers are not equal, the
 string can be driven towards the chiral limit.
 This will happen particularly near a cusp, where the string is 
contracted by a large factor, 
resulting in a much higher density of charge carriers than elsewhere along
the string and in a great enhancement of the scattering rate. One can expect
therefore that the current will be very nearly chiral in the vicinity of 
cusps. In the case of electrically charged currents, the electromagnetic 
radiation from oscillating loops is dominated by powerful bursts emitted from
near-cusp regions\cite{Vilenkin87,Spergel87}. One can therefore use the
solution for strings with chiral currents to estimate the radiation power
from such loops.

To avoid confusion, we note that cusps were originally 
defined\cite{Turok84} as points of infinite contraction, where the string
momentarily reaches the speed of light. Strictly speaking, such cusps can
be formed only on idealized Nambu-Goto strings. For realistic strings, the
cusp development is truncated either by the annihilation of overlapping
string segments at the tip of the cusp\cite{Branden87,jjkdo98.0,jjkdo98.1} 
or for superconducting strings, by the back reaction of charge
carriers or of the electromagnetic radiation. However,
unless the string current is very large, so that the energy
 of the charge carriers is comparable to that of the string itself, the 
truncation occurs at a very large Lorentz factor and the string exhibits
cusp-like behavior. Below we shall use the word ``cusps'' to refer to such 
ultra-relativistic string segments.

Carter and Peter derived their solution\cite{Carter99-1} using Carter's
formalism\cite{Carter89-1,Carter89-2} which is not familiar to most 
cosmologists.
In view of the importance of this result, in the present paper we shall
give an alternative, elementary derivation. We shall also give an explicit 
solution for the initial value problem and discuss several physical 
implications of the result.

We begin in Section II by reviewing the well-known solution of string
equations of motion in the case of non-superconducting Nambu-Goto
strings.  Our derivation of the solution for chiral superconducting
strings is presented in Section III. In Section IV, we show how the
motion of such strings can be found from given initial
conditions. Some physical implications of the solution are discussed
in Section V, where we show that the motion of string loops with
chiral currents is strictly periodic and calculate the maximum Lorentz
factor reached at the cusp. In Section VI we argue that the chiral
string solution can be used as an approximate description of strings
with small non-chiral currents. In Section VII we use this approximate
description to estimate the electromagnetic radiation from oscillating
loops with charged currents.

After this paper was submitted we learned about independent work by Davis 
et al.\cite{Davis00-1} which gives a derivation of the chiral string solution
similar to ours.

\section{Nambu-Goto strings}
We first review the equations of motion and their solution in the case
of non-superconducting strings.  We will use similar techniques in the
next section to solve the equations of motion for superconducting
strings, in the case that the current is chiral.

For an infinitely thin non-superconducting relativistic
string, the flat spacetime equations of motion are\cite{Alexbook}
\be
\partial_a \left[ \sqrt{- \gamma} \,  \gamma^{ab} \, x^{\nu}_{, b}
\right] = 0
\ee
where $a$ and $b$ take the values 0 and 1 (denoting the worldsheet
 coordinates $\tau$ and $\sigma$ respectively), $x^{\nu}
(\sigma, \tau)$ is the position of the
string, and $\gamma_{ab}$ is the induced metric on the worldsheet,
\be\label{eqn:gammax}
\gamma_{ab}= x^\mu_{, a} x_{\mu, b}
\ee
and $\gamma =\det(\gamma_{ab})$.  We are free to choose a particular
parameterization of the worldsheet, i.e. gauge condition.  In this case,
it is convenient to use the conformal gauge, namely
\be\label{eqn:conformal}
\gamma_{ab} = \Omega (\sigma,\tau)\, \eta_{ab}
\ee
where $\eta_{ab}$ is the two-dimensional Minkowski metric.  Then
$\sqrt{-\gamma} =\Omega$ and thus $\sqrt{-\gamma}\gamma^{ab}
=\eta^{ab}$.  In this gauge the equation of motion for the string is
just the two dimensional wave equation,
\be\label{eqn:wave}
 x''^{\nu} - \ddot x^{\nu} = 0\,,
\ee
where $\dot x$ denotes $\partial x/\partial\tau$ and $x'$ denotes
$\partial x/\partial\sigma$.

It can be seen that the constraints written above in Eq.\
(\ref{eqn:conformal}), do not fix completely our gauge, and we can
choose the condition
\be
x^0 = \tau
\ee
so the equations of motion become
\be
\bx'' - \ddot\bx =0\,.
\ee
The general solution has the form
\be\label{eqn:ab} 
\bx = {1\over 2}[\ba(\sigma - \tau) + \bb(\sigma + \tau)]\,.
\ee
In order for the metric to have the form of Eq.\
(\ref{eqn:conformal}), we must impose the conditions
\be
|\ba'|^2 = |\bb'|^2 = 1\,.
\ee

\section{Chiral superconducting strings}

We consider a superconducting string with a neutral
current (i.e., one not coupled to the electromagnetic field).
We can describe the current via an auxiliary scalar field $\phi$, in
terms of which the conserved worldsheet current is
\be\label{eqn:current}
J^a = {1 \over{\sqrt{-\gamma}}} \,\epsilon^{ab} \, \phi_{,b} \, .
\ee
The current is chiral if $\phi_{,a}$ is a null worldsheet vector, i.e.,
\be\label{eqn:chirality} 
\gamma^{ab}\phi_{, b}\phi_{, a} = 0\,,
\ee
in which case $J_{a}J^{a} = 0$.  In this case, the equations of motion
can be written \cite{Witten85,Vilenkin87,Carter99-1}
\blea
\partial_a \left( {\cal T}^{ab} x^{\nu}_{, b} \right) &=& 0\\
\partial_a\left(\sqrt{-\gamma}\gamma^{ab}\phi_{,b}\right) &=& 0
\label{eqn:phimotion}
\elea
where
\be \label{eqn:Tdef}
{\cal T}^{ab} = \sqrt{- \gamma} \, \left(\mu \gamma^{ab} +
 \theta^{ab}\right)\,,
\ee
$\mu$ is the energy per unit length of the string, 
and $\theta^{ab}$ is the worldsheet energy-momentum tensor of the charge
carriers,
\be\label{eqn:thetadef}
\theta^{ab} = \gamma^{ac}\gamma^{bd}\phi_{,c} \, \phi_{,d}
\ee
in the chiral case.

As above, we would like to
have a gauge in which the ${\cal T}^{ab}$ has the form
\be\label{eqn:Tform}
{\cal T}^{a b} =\mu \, \eta^{a b}\,.
\ee 
If we can accomplish this, the equation of
motion will be the wave equation, Eq.\ (\ref{eqn:wave}), we can
choose $x^0 =\tau$, and the general solution will be given by
Eq.\ (\ref{eqn:ab}), as before.
However, we note that since ${\cal T}$ is a $2\times 2$ matrix,
\bea
\det {\cal T} &=& (-\gamma)\det\left(\mu \gamma^{ab} + \theta^{ab}\right)
\nonumber\\
&=&-\det\left[\gamma_{a b}\left(\mu \gamma^{bc} + \theta^{bc}\right)\right]
 = -\det\left(\mu\delta^c_a+\theta^c_a\right)
\eea
which is gauge-invariant.  Since the matrices are $2\times 2$, the
determinant is easily expanded,
\be\label{eqn:detexp}
\det {\cal T} = -\mu^2-\mu\Tr\theta^c_a -\det\theta^c_a\,.
\ee
Since $\theta$ is traceless, $\det {\cal T}$ can only be $-\mu^2$ as required
by Eq. (\ref{eqn:Tform}), if $\det\theta^c_a = 0$.  But for (and only
for) a chiral current, from Eq.\ (\ref{eqn:thetadef}),
\be
\theta^c_a =\phi^{, c}\phi_{, a}\,,
\ee
which is the outer product of two vectors, and thus has vanishing determinant.

Thus for a chiral current there is the possibility that
Eq.\ (\ref{eqn:Tform}) can be satisfied.  In that case, from Eqs.\
(\ref{eqn:Tform}) and (\ref{eqn:Tdef}) we see that
\be
\sqrt{-\gamma}\gamma^{a b}\phi_{,b}
={1\over\mu}\left[{\cal T}^{ab} -\sqrt{-\gamma} \theta^{ab}\right]\phi_{, b}
=\eta^{ab}\phi_{,b}
\ee
and so Eq.\ (\ref{eqn:phimotion}) becomes the wave equation,
\be\label{eqn:phiwave}
\ddot\phi -\phi'' = 0\,.
\ee

We now contract Eq.\ (\ref{eqn:Tform}) with $\phi_{,a}\phi_{,b}$. Since 
the current is chiral, $\gamma^{ab}\phi_{,a}\phi_{,b} = 0$, and
using Eq.\ (\ref{eqn:thetadef}), $\theta^{ab}\phi_{,a}\phi_{,b} = 0$.
Thus to satisfy Eq.\ (\ref{eqn:Tform}), we must have
$\eta^{ab}\phi_{,a}\phi_{,b} = 0$, or $\dot\phi^2 =\phi'^2$.  Without 
loss of generality we take the solution of the form
\be\label{eqn:phi}
\phi (\sigma,\tau) = F (\sigma +\tau)\,,
\ee
which also satisfies Eq.\ (\ref{eqn:phiwave}).

Using Eq.\ (\ref{eqn:phi}), the condition for chirality,
Eq.\ (\ref{eqn:chirality}), becomes
\be\label{eqn:gammaup}
\gamma^{00} +\gamma^{11} +2\gamma^{01} = 0\,.
\ee
Since $\gamma^{ab}$ is the inverse of the $2 \times 2$ matrix $\gamma_{ab}$ we have
\be
\gamma^{ab} = {1\over {\gamma}}
 \left( \begin{array}{cc}
\gamma_{11}  & {-\gamma_{01}} \\
{-\gamma_{01}} &  \gamma_{00} 
\end{array}
\right)\,,
\ee
so Eq.\ (\ref{eqn:gammaup}) implies
\be\label{eqn:gammadown}
\gamma_{00} +\gamma_{11} -2\gamma_{01} = 0\,.
\ee
Using Eq.\ (\ref{eqn:gammax}), this becomes
\be
0=\dot x^\mu \dot x_\mu + x'^\mu x'_\mu - 2 \dot x^\mu x'_\mu = 
(\dot x^\mu- x'^\mu)(\dot x_\mu-x'_\mu)
\ee
which means that $\dot x^\mu- x'^\mu = (1,- \ba')$ is a null 4-vector, or that
\be\label{eqn:Tconstraint}
|\ba' | = 1\,.
\ee
To solve the rest of the problem, we define a matrix
\be\label{eqn:Sdef}
{\cal S}_{ab} ={1\over\sqrt{-\gamma}}\left(\mu\gamma_{ab} -\theta_{ab}\right)\,.
\ee
From Eqs.\ (\ref{eqn:chirality}) and (\ref{eqn:thetadef}),
$\theta_{ab}\theta^{bc} = 0$, and since $\gamma^{ab}$ and $\theta^{ab}$ are
symmetrical,
\be
{\cal S}_{ab} {\cal T}^{bc} =\mu^2 \, \delta^c_a
\ee
and consequently if ${\cal T}$ has the form of Eq.\ (\ref{eqn:Tform}) we have
that
\be\label{eqn:Sval}
{\cal S}_{ab} =\mu \eta_{a b}\,.
\ee
Now, comparing the $01$ component of Eqs.\ (\ref{eqn:Sdef}) and
(\ref{eqn:Sval}) gives

\be\label{eqn:F}
\mu \gamma_{01} =\theta_{01} = F'^2 \, .
\ee
The metric component is
\be\label{eqn:b1}
\gamma_{01} =\dot x^\mu x'_\mu ={1\over 4}\left(|\ba '|^2- |\bb' |^2\right)\,.
\ee
From Eqs.\ (\ref{eqn:b1}) and (\ref{eqn:Tconstraint}), we find
\be\label{eqn:Sconstraint}
1- |\bb' |^2 = {{4 \, F'^2}\over {\mu}}
\ee
Since the determinant of ${\cal T}$ and thus of ${\cal S}$ is fixed, it remains only
to show that one more component of Eq.\ (\ref{eqn:Sval}) is
satisfied.  For example, a sufficient condition is that
\be\label{eqn:lastconstraint}
\mu \gamma_{00} -\theta_{00} =\mu \sqrt{-\gamma}\,.
\ee
Using Eq.\ (\ref{eqn:gammadown}) it is easy to show that
\be
\sqrt{-\gamma}  = {{1}\over 2} \, (\gamma_{00} -\gamma_{11}) \, ,
\ee
and Eq.\ (\ref{eqn:lastconstraint}) becomes
\be
\mu\gamma_{00} - F'^2 = {{\mu}\over 2} (\gamma_{00} -\gamma_{11})
\ee
so
\be
{{\mu}\over 2} (\gamma_{00} +\gamma_{11}) = F'^2
\ee
which is satisfied using Eq.\ (\ref{eqn:gammadown}) and 
Eq.\ (\ref{eqn:F}).

Thus, Eq.\ (\ref{eqn:ab}) with the constraints given by Eqs.\
(\ref{eqn:Tconstraint}) and (\ref{eqn:Sconstraint}) are a general
solution to the equations of motion for a superconducting string with
a chiral current. This solution is the same as that found by 
Carter and Peter\cite{Carter99-1}, except they did not explicitly 
specify the relation (\ref{eqn:Sconstraint}) between $\bb'$ and $F'$, 
but gave instead the inequality $|\bb' |^2 < 1$.

Note that in this gauge $\sigma$ parameterizes the total energy on the
string.  The energy in a region is
\be
E = \int{d^3 x \, T^0_0}\,,
\ee
where $T^{\mu}_{\nu}$ is given by
\be
T^{\mu}_{\nu}(x) = \int{d\sigma \, d\tau\, {\cal T}^{a b} x^{\mu}_{,a} \,  x_{\nu ,b} \, \delta^4[x-x(\sigma,\tau)]}.
\ee
With $x^0 = \tau$, and using  Eq.\ (\ref{eqn:Tform}), the energy is
\be\label{eqn:eperlength}
E = \mu \, \int{d\sigma}\, = \mu \, \int{dl {1\over |\bx'|}}
\ee
which means that with this parameterization the energy on the string
is $\mu \, \Delta \sigma$.

\section{Finding the solution from initial conditions.}
We would now like to find the evolution of a string with a chiral
current from given initial conditions.  We suppose that we are given
the position of the string at some time $t_0$, as a function $\bx(l)$
parameterized by arc length in the laboratory frame, with $l$
increasing in the opposite direction to the current flow.  We also need the
initial charge and current as the values of the auxiliary scalar field
$\phi(l)$, and the perpendicular component of the string motion,
$\xdot_\perp (l) $.  Motion parallel to the string direction is
dependent on the choice of parameter and has no physical meaning.
From these conditions, we want to find the functions $\ba$ and $\bb$.

The first step is to reparameterize everything in terms of $\sigma$.
For a stationary string, the linear energy density of the string
itself is just $\mu$, and the energy due to the current is
$(d\phi/dl)^2$.  Boosting the string in a transverse direction just
gives the Lorentz factor $\Gamma = 1/\sqrt{1- |\xdot_\perp |^2}$, so
\be\label{eqn:dEdl}
{dE\over dl} =\Gamma \left[\mu +{\left({d\phi\over
dl}\right)}^2\right]
\ee
and thus
\be\label{eqn:dsdl}
{d\sigma\over dl} =\Gamma \left[1 +{1\over\mu}{\left({d\phi\over
dl}\right)}^2\right]\,.
\ee
Using Eq.\ (\ref{eqn:dsdl}) we can change parameters from $l$ to
$\sigma$.

Now we need to determine the full form of $\xdot$. We observe that
$\xdot\cdot\bx' = (|\bb' |^2-|\ba' |^2)/4 = -\phi'^2/{\mu}$, so we can write
\be
\xdot =\xdot_\perp -{\phi'^2\bx'\over{\mu \, |\bx' |^2}}\,.
\ee
Then
\blea
\ba' &=&\bx' -\xdot\\
\bb' &=&\bx' +\xdot
\elea
and Eq.\ (\ref{eqn:ab}) gives a complete solution for the future
evolution of the string.

\section{Chiral String Dynamics}
We have obtained the analytic general solution for superconducting strings with
chiral currents. For a string with a current determined from,
\be\label{eqn:general-1}
\phi (\sigma,\tau) = F (\sigma +\tau)\,,
\ee
the general solution for the string is given by
\begin{mathletters}\label{eqn:general-2}
\begin{eqnarray}
x^0 & = & \tau \\
\bx & = & {1\over 2}[\ba(\sigma - \tau) + \bb(\sigma + \tau)]\,, 
\end{eqnarray}
\end{mathletters}
with the following constraints for the otherwise arbitrary functions $\ba'$ and $\bb'$,
\begin{mathletters}\label{eqn:general-3}
\begin{eqnarray}
|\ba'|^2 &=& 1 \\
|\bb'|^2 &=& 1 -  {{4 \, F'^2}\over {\mu}}\,.
\end{eqnarray}
\end{mathletters}

Note that $F'= d\phi/d\sigma$ is in general not the same as the
physical current, which goes as $d\phi/dl$.  Using
Eq.\ (\ref{eqn:dsdl}) we can write
\be
{d\phi\over dl} = \Gamma \left[1 +{1\over\mu}{\left({d\phi\over
dl}\right)}^2\right] F'
\ee
so
\be\label{eqn:Fmax}
F' = {\mu(d\phi/dl)\over [\mu +(d\phi/dl)^2] \Gamma}\,.
\ee
Thus $F'$ increases with $d\phi/dl$ for small currents, but it
reaches a maximum, $F' =\sqrt{\mu}/(2\Gamma)$ when $d\phi/dl
=\sqrt{\mu}$.  For larger values of $d\phi/dl$, $F'$ decreases again.
This happens because $F'$ measures the winding number per unit energy,
and the current contribution to the energy density goes as $(d\phi/dl)^2$.

We now show several interesting consequences that we can extract from
the result.  First of all, we see that there are arbitrarily shaped static
solutions for the case in which the current satisfies $4\, {F'}^2 /\mu
= 1$. In this case $|\bb'|=0$, so the position of the string, up to a 
constant vector, is given by
\be
\bx =  {1\over 2}[\ba(\sigma - \tau)]
\ee
so the set of points traced by the string does not depend on
time. These are vortons of any shape, which could have important
cosmological consequences.

Another somewhat unexpected consequence of the exact solution is that the 
motion of a loop with a chiral current is strictly periodic in its rest frame.
The period is $T = E/(2\mu)$, where E is the total energy of the loop in that
frame.

We also see that chiral strings do not have true
cusps. In fact, we can easily calculate the Lorentz factor
that these strings can reach,
\be
\Gamma = {1\over {\sqrt{ 1 - {\xdot_\perp}^2}}} =
\sqrt{1 + \left( {{|\bb'| \sin \theta}\over {1 + |\bb'| \cos \theta}}\right)^2 }\,,
\ee
where $\theta$ is the angle between $\bf a'$ and $\bf b'$. This expression
has its maximum at $\cos \theta = - |\bb'|$, and the maximum 
Lorentz factor is
\be\label{eqn:Gammamax}
\Gamma_{\rm max} = {{\sqrt \mu}\over {2 |F'|}} \,.
\ee

This result can be seen directly from Eq.\ (\ref{eqn:Fmax}).  At this
point, the energy density in the string itself and that in the charge
carriers (see Eq.\ (\ref{eqn:dEdl})), are equal.

However, this is not the maximum concentration of energy per unit
length that can be achieved.  Using Eq.\ (\ref{eqn:eperlength}) we see
that the maximum energy density corresponds to the minimum of $|\bx
'|$, i.e. $\theta =\pi$.  There,
\be
|\bx'|_{\rm min} = {1\over 2}\left({ 1 - |\bb'|}\right)
\ee
and so
\be
\left. {dE\over dl}\right |_{\rm max} =
{2\mu\over 1- |\bb' |} = {2\mu\over 1-\sqrt{1-4F'^2/\mu}}\,.
\ee
At this point $\ba'$ and $\bb'$ are antiparallel, and thus so are
$\xdot$ and $\bx'$.  Thus $\xdot_\perp = 0$ and the string is not
moving, so this is also the point of maximum energy density in the
local rest frame.

\section{Strings with a small non-chiral current}

The Carter-Peter solution for chiral strings,
Eqs.\ (\ref{eqn:general-1})--(\ref{eqn:general-3}) can also be
used as an approximate description of strings having both right- and
left-moving charge carriers in the case when the current is
sufficiently small,
\be
{\dot\phi}^2 , ~ {\phi'}^2 \ll \mu.
\label{smallJ}
\ee
The contribution of charge carriers to the worldsheet energy momentum
tensor is then suppressed compared to that of the string itself
by a factor $\sim \Gamma^2\phi'^2$, where $\Gamma$ is the Lorentz
factor of the string.  (We assume for simplicity that ${\dot\phi}^2
\sim {\phi'}^2$.)  The effect of charge carriers on the string
dynamics will therefore be negligible, except in the vicinity of cusps
where $\Gamma$ can be very large.

The string current is also greatly enhanced in near-cusp regions.  For
strings with bosonic superconductivity, large currents can be unstable
with respect to quenching \cite{Davis88-1} and to quantum 
tunneling\cite{Witten85,Davis88-3,Zhang87,jjkdoav00-2}.
A large charge density can also destabilize the condensate, resulting
in ejection of charge carriers.  These effects disappear for a chiral
current and tend to drive the string towards the chiral limit
\cite{Davis88-2,Martins98}.  For strings with fermionic
superconductivity, the growth of the current leads to enhanced
 scattering of left- and
right-moving charge carriers off the string.  The left- and
right-moving currents are typically not equal, so the scatterings
suppress the minority charge carriers, leaving the string with nearly
chiral current in the near-cusp region\cite{Spergel87}.

We thus have a situation where away from the cusps the current is
non-chiral but small and its effect on the string dynamics is
negligible, while near the cusps the current is large and nearly chiral.
Hence, both near cusps and away from cusps the string dynamics can be
approximately described by the chiral string solution,
Eqs.\ (\ref{eqn:general-1})--(\ref{eqn:general-3}).

We shall now derive a quantitative criterion for this approximate
description to be accurate.
Near a cusp, the string gets contracted by a large factor, its rest
energy being turned into kinetic energy.  The density of charge
carriers and the current are enhanced by the same factor.  The
contraction factor increases as we approach the tip of the cusp.  
The invariant length of string which attains Lorentz factor at least
$\Gamma$ is $\sim L/\Gamma$.  Since this Lorentz factor is obtained by
compressing the string, the physical length of this region of string
is
\be
\Delta L_\Gamma\sim L/\Gamma^2\,,
\ee
and since the physical string motion is perpendicular to the string,
this is also the length of string in its rest frame.  Let $J
=\sqrt{|J_\mu J^\mu|}$ and let $J_0$ be the value of $J$ far from the cusp.
Since the string has been compressed by factor $\Gamma$, the current
is increased by the same factor, so it becomes
\be
J_\Gamma \sim \Gamma J_0.  
\label{Jgamma}
\ee
These values are sustained for a time
interval (again in the rest frame)
\be
\Delta t_\Gamma\sim \Delta L_\Gamma\sim L/\Gamma^2\,.
\label{tgamma}
\ee

We now have to check whether or not this time interval is sufficient
to suppress the minority charge carriers.  The answer to this depends
on the scattering rate of left- and right-movers and is therefore
model-dependent.  As an illustration, let us suppose that we have
fermionic charge carriers with large masses off the string but which
can scatter inelastically by exchange of a GUT-scale gauge boson into
light particles not bound to the string.  Models of this sort have
been studied in detail by Barr and Matheson \cite{Barr87-1,Barr87-2}.
Their analysis indicates that the time it takes for the current to
become nearly chiral is (in the rest frame of the string)
\be
\tau\sim M_X^4/J_\Gamma^5,
\label{tau}
\ee
where, $M_X$ is the GUT scale.  The current will be nearly reduced to one
chiral component during a single cusp occurrence, provided that this
time is shorter than $\Delta t_\Gamma$.  Using Eqs. (\ref{Jgamma}),
(\ref{tgamma}) and (\ref{tau}) we find that this condition is
satisfied when the Lorentz factor exceeds a certain value $\Gamma_*$,
\be
\Gamma \gg\Gamma_*\sim\left({M_X^4\over{LJ_0^5}}\right)^{1/3}.
\ee

On the other hand, charge carriers have a substantial effect on the
string dynamics when the Lorentz factor becomes comparable to
$\Gamma_{\rm max}$ from Eq.\ (\ref{eqn:Gammamax}).  Thus, we expect the approximate chiral
string description to be accurate when $\Gamma_*\ll\Gamma_{\rm max}$, or
\be\label{eqn:currentbound}
{J_0\over{\sqrt{\mu}}}\gg\left({M_X^2\over\mu}\right)^{5/4}(M_X
L)^{-1/2}.
\ee

This condition is extremely weak for macroscopic strings.  For
example, for superheavy strings with $\mu\sim M_X^2$, $\sqrt{M_XL}$ is
the maximum Lorentz factor which can be reached before overlap of the
string cores causes the cusp to be truncated
\cite{jjkdo98.0,jjkdo98.1}.  Thus if Eq.\ (\ref{eqn:currentbound}) is
violated, then neither $\Gamma_*$ nor $\Gamma_{\rm max}$ can be
reached, and the current has negligible effect on motion of the string.

\section{Electromagnetic radiation power from oscillating loops}

If the string current is coupled to electromagnetism,  then
oscillating current-carrying loops emit electromagnetic radiation.
Calculations disregarding the effect of charge carriers and of the
electromagnetic back-reaction on the motion of the loop give an
infinite radiation power.  The divergence can be
attributed to the infinite Lorentz factor reached at the cusp of a
Nambu-Goto string.  If the cusp is truncated at a maximum Lorentz
factor $\Gamma_{\rm max}$, the power can be estimated as \cite{Vilenkin87,Spergel87}
\be\label{eqn:P}
P\sim 30 j^2\Gamma_{\rm max}.
\ee
Here, $j \sim  q J_0$ is the electric current away from the cusp, $q\sim0.1$
is the effective charge of the charge-carrying field, and the coefficient comes
from numerical calculations. 

Later in this Section, we shall argue that the electromagnetic back-reaction
in the near-cusp region is subdominant compared to the effect of the charge carriers and can therefore be neglected.

As discussed in the preceding Section, for sufficiently small currents the 
charge-carrier back-reaction is negligible away from the cusps, while near
the cusps the current tends to be chiral and the solution
Eqs.\ (\ref{eqn:general-1})--(\ref{eqn:general-3}) can be used. Using this solution
 we can determine $\Gamma_{\rm max}$ and therefore we can give an estimate for the electromagnetic power.
Substituting Eq.\ (\ref{eqn:Gammamax}) in Eq.\ (\ref{eqn:P}) we obtain
\be\label{eqn:P1}
P\sim qj\sqrt{\mu}\,.
\ee
This result is confirmed by more accurate calculations in
\cite{jjkdo00.2}.

Our neglect of the electromagnetic back-reaction can be justified as
follows.  The energy emitted in a single cusp event is $\Delta
E_{em}\sim PL$, where $L$ is the length of the loop.  The total energy
of the string segment in which the maximum Lorentz factor is reached
(and which is responsible for most of the radiation) is $\Delta
E_s\sim \mu L/\Gamma_{\rm max}$, and the energy of the charge carriers
in that region is $\Delta E_J\sim \Delta E_s$.  Using Eqs.\
(\ref{eqn:Gammamax}, \ref{eqn:P1}), we find $\Delta
E_{em}/\Delta E_J \sim q^2 \ll 1$.  This suggests that the effect of
electromagnetic back-reaction on the motion of the string is much
smaller than that of the charge carriers.

\section{Acknowledgments}

We would like to thank Anne Davis and Xavier Siemens for helpful
conversations. This work was supported in part by the National Science
Foundation.


\end{document}